# Conceptual Knowledge Relationship Model of Enterprise Architecture and Top Management Roles

**Hawa Ahmad**
Department of Accounting
Kulliyyah of Economics and Management Sciences
International Islamic University Malaysia (IIUM)
Malaysia
Email: hawahmad@iium.edu.my

**Aishah Ahmad[1] Nik Hazliana Nik Abdullah Hafaz[2]**
Faculty of Computer and Mathematical Sciences
Universiti Teknologi MARA (UiTM)
Malaysia
Email: aishah582@melaka.uitm.edu.my[1]
Email: nikhazliana@gmail.com[2]

## Abstract

Enterprise Architecture (EA) continues to gain global recognition as a management tool that would improve the organization's performance. The literature review reveals that the alignment of EA with that of the business strategy was not achieved due to inadequate EA knowledge at Top Management (TM) level. This study aims to gauge the level of EA knowledge required at TM level to enable the creation of EA that would enable the alignment of strategic business vision with that of EA. A semi-structured interview was conducted with several personnel in an organization. Data were analyzed using the constant comparative method. This study identified that the TM need to have understanding to two sets of EA knowledge, viz, business process management (BPM) and technology knowledge. A conceptual knowledge relationship model is proposed through synthesizing the roles of EA and TM in supporting the TM in identifying the EA knowledge required. The findings gave contribution to academicians and practitioners with knowledge of how to improve EA assimilation and a structured roadmap for understanding EA to achieve high business value.

## 1   Introduction

Enterprise Architecture (EA) is an accepted method in addressing the challenge to align between business and information technology through a holistic view (Ross et al. 2006; Saha 2006; Open Group 2006; Tucker and Aron 2005: USA Federal CIO Council 2001). The numerous benefits and promises of EA have attracted many organizations worldwide to extensively invest in EA initiatives. EA has generally evolved into a well-accepted discipline and its importance is considered to be growing (Schekkerman 2005). It also acts as a management tool that could assist in the integration (Lemmeti and Pekkola 2012).

However, studies reported that more than 66% of EA initiatives failed in reaching organization's anticipated goal, serious budget overruns with disappointing performance results (Kappelman 2009; Broer 2005; Markus and Tanis 2000). EA adoption is reported to be a challenging and complex organizational learning and change management process ( van der Raadt et al. 2007; Zachmann 2003; Armour et al. 1999). Despite the attempts to lower the degree of complexity of EA and the various developments of EA, many organizations are still experiencing failures with their EA (Gaver 2010). Markus and Tanis (2000) claim that while a rapid and smooth adoption might reveal initial success, implementation quality can result in underutilization of EA. Graver (2010) added that an effective EA implementation, although a necessary prerequisite does not provide a sufficient condition for an organization to fully benefit from EA.

In order to generate significant business value, EA should be integrated and embedded in the corporate value chain before it can generate business value (Delone and McLean 2002). Existing evidence suggests that the potential value of such complex innovations (Perko 2008) can only be realized through persistent and successful assimilation within the organization (Armstrong and Sambamurthy 1999). In contrast to the implementation or adoption, assimilation involves actual usage of the new innovation to the extent the innovation is utilized and routinized or becomes part of the culture of the operational structure of the organization (Perko 2008; Purvis et al. 2001).

This study builds on the work of Rafidah et al. (2007) to advance an understanding of the EA practices in Malaysia. Their research revealed that these organisations practice EA although incomplete or inadequate. This situation is due to lack of awareness of the need for the alignment between the strategic objectives of the business and that of EA at TM level. The study was limited by its methodology. An earlier exploratory study of EA practices in Malaysia by Zulkhairi et al (2006) also identified the EA knowledge barrier among TM. However, their study was an exploratory, no approaches were used to establish better understanding of EA in order to further stimulate the assimilation of EA.

According to Jasperson et al. (2005), limited knowledge can hamper a person's ability to understand and exploit system capabilities available therefore limit the system assimilation. Similarly Perko (2008) claims that inadequate knowledge may limit organizations to better understand their EA and business process environment. As a result, failing to utilize the full extent of EA would eventually affect the EA assimilation process. In fact, knowledge barriers and knowledge burden are one of the factors known from the literature as affecting EA assimilation (Fichman and Kemerer 1999; Perko 2008). Therefore, the effective use of the new technology can be realized when the knowledge barriers and the knowledge burden are lowered or minimized (Perko 2008; Weil and Ross 2003).

The aim of this study is to determine the EA knowledge required for TM so that they can provide a comprehensive support on EA assimilation.

The objectives of this study are:

- To identify the roles of EA in an organization.
- To identify the roles of TM in an organization to support the EA assimilation.
- To synthesize the fit between the relevant roles in supporting the TM to identify the EA knowledge required.

The remaining part of this paper is structured as follows. The next section will provide a review of relevant literature on the roles of EA and TM. This is followed by research method, analysis and findings. The conclusion, limitations and suggestions for future research are then discussed.

## 2   LITERATURE REVIEW

This review focus on the particular roles played by EA and the level of involvement requirements of TM role players to advance the EA knowledge and obtain support for its effective assimilation at the enterprise level.

### 2.1   Enterprise Architecture Roles

Enterprise Architecture (EA) acts as a holistic blueprint that provides a long-term view of an organization's processes, systems, and technologies (Ross et al. 2006; Saha 2006). EA provides an implementation roadmap or a transition strategy to guide the implementation of the target blueprint. According to Lankhorst (2005) a target blueprint structures the overall solution in business and information, information systems and technical infrastructure layers. An implementation roadmap describes how the vision of the enterprise can be achieved by going from the "as-is" state through a set of intermediate states to the "to-be" state of the enterprise (Pulkkinen et al. 2007).

According to Armour et al. (2005), an EA identifies the main components of the organization, its information systems, the ways in which these components work together in order to achieve defined business objectives, and the way in which the information systems support the business processes of the organization (Armour et al. 2005). The components include staff, business processes, technology, information, financial and other resources, etc. (Armour et al. 1999). In a simpler manner, according to Kappelman (2010), EA is viewed as a systematic application of architectural fundamentals to manage the complexity of enterprises (Kappelman 2010).

Originally, EA is a method for IT management and system architectures (Zachman and Sowa 1992; Zachman 1987). However, in recent years, EA has grown into a general management discipline, promoting several areas of planning, consolidating, and aligning strategic initiatives, management programs, capital planning, business processes, and IT assets (Jensen 2010). EA from a holistic point of view as "a complete expression of the enterprise; a master plan which acts as a collaboration force between aspects of business planning such as goals, visions, strategies and governance principles; aspects of business operations such as business terms, organization structures, processes and data; aspects of automation such as information systems and databases; and the enabling technological infrastructure of the business such as computers, operating systems and networks" (Schekkerman 2005).

The growing interest of the EA concept was confirmed by several studies conducted in the 5 year period between 2003 and 2008 (i.e Institute for Enterprise Architecture Developments, Jonathan Broer for the Rotterdam University, Gartner). The results demonstrate that EA is applied in more and more organizations in the world, and their numbers increase in countries like South Korea, Japan, China, India, Iran, Russia (Tamm et al. 2011; Dankova 2009).

EA is implemented mostly by large organizations; however, there has been an increased in numbers in small and medium enterprises reaching positively to the concept. In addition, studies also show that EA is deployed mostly by government agencies in the field of industry, energy and municipal services, transport, financial services, as well as healthcare.

## 2.2 Top Management Roles

This study takes on a stance on the call for TM support with a clear assumption that top managers' understanding and effective involvement have significant impact on realizing EA practices (Ross et al. 2006).

Top management (TM) is a team of individuals at the highest level of organizational management who have the day to day responsibilities of managing an organization. Top managers, for example, have a great impact on the decision making and ultimately on the outcomes achieved the organization (Hambrik 2004; Hambrick and Mason 1984). Besides discharging specific responsibilities allocated to top managers such as those related to functions like marketing, production, finance or personnel management, in general top managers play strategic leadership, operating, and many other roles in an organization (Akhouri 2002; Robbins 1998). The performance of the organisation is influenced by how well the multiple roles are played. A critical element of strategic leadership and the effective implementation of strategy is the ability to manage the organization's resource portfolio. This includes integrating resources to create capabilities and leveraging those capabilities through strategies to build competitive advantages (Hite et al. 2007).

The TM championship has been consistently identified to be one of the most critical factors, both in IS implementation and innovation studies, such as EA (Purvis et al. 2001). It refers to the extent that TM supports, directly and indirectly, and commits to the continuous use of IS projects. TM involvement and their sustained support throughout the phases of the project to help ensure a smooth change management and mobilizing commitment of other stakeholders (Somers and Nelson 2004). The degree of TM commitment is a crucial element in shaping the EA functions setup and to ensure sufficient resources. Insufficient TM commitment has been an important issue. In addition, Ross et al. (2006) added in EA practice, the involvement of TM does not stop at the planning stage, instead they should be able to demonstrate the understanding of their organization's EA and provide oversight on architecture initiatives (Ross et al. 2006). Research has even shown that TM support is the most predictive factor of any IS project success (Somers and Nelson 2004).

According to Seppanen et al. (2009) "business and IT managers are primarily responsible for creating a favorable atmosphere that is required in ensuring that the architectural process is granted enough time, money and other resources". Involving experts with important knowledge is essential: "We've seen projects fail because key people were put on them part-time". Therefore the leadership needs to be committed to and communicate passion and excitement for EA (Lange and Mendling 2011). A study by Fichman and Kemerer (1999) however mentioned that failure to recognize an innovation's value may be a reflection of TM's lack of knowledge of its usage.

## 2.3 Enterprise Architecture Assimilation

Enterprise Architecture (EA) challenges and failures have been reported in the literature. It was reported that there is an increasing awareness of EA that has not been working well in many organizations, both in industry or in government (Graver 2010). In addition, there is a lack of consensus on EA concepts, terminologies, goals, approaches, techniques, and outcomes (Graver 2010). Aziz et al. (2005) who carried out a research with Infosys Ltd stated that EA hardly ever fails because of inadequate content. He added that the challenges usually arise around how to link the EA efforts into the overall enterprise processes, and how to leverage them as assets used regularly by a variety of stakeholders.

Several researchers claimed that the potential value of an IS innovation can only be fully realized when they are extensively assimilated in an organization, adding that a successful IS implementation does not automatically lead to continued use of the system by the organization (Liang and Zhu 2007). Quoting Cooprider and Victor (1993), Makiya and Lyytinen (2011) mentioned that studies also found evidence that high levels of information system (IS) and business domain knowledge can also enhance IS assimilation in organizations. Similarly, it is posit that positive value recognition of EA among top executives is critical to EA assimilation (Makiya 2011). As such, the firm needs to deeply understand the system's technology and capabilities, and to integrate it into the business functions in order to efficiently assimilate the system (Chatterjee et al. 2002).

The key aspect is that assimilation as a process of 'infusion' is a lengthy process of integrating and institutionalizing the innovation into the operational and social structure of the organization. Fichman and Kemerer (1999) introduced innovation an assimilation model that identifies seven stages which incorporate the innovation from adoption into full scope assimilation: contact; awareness; understanding; trial Use/Training; adoption; Institutionalization and finally Internalization. While Adoption refers to the decision about using or not the technology (Chatterjee et al. 2002; Purvis et al. 2001; Armstrong and Sambamurthy 1999), Assimilation on the other hand can be classified as a series of stages that follow the organization's initial formal adoption and lead to a widely accepted and extensive deployment of the system where it becomes a custom as well as a significant component of the organization's activities (Fichman 2000).

The review of the literature showed that not only is there a lack of understanding for both EA practice and its assimilation, studies also showed lack of strong executive sponsorship and ongoing leadership as a major problem in EA development efforts. Despite the huge claim for TM, most studies within the EA literature did not study into details on the assimilation. Typically EA researchers acknowledge TM's importance, but they did not take the discussion further and therefore surprisingly little in-depth research on the TM support of EA has been published.

The first thing which TM must realize from the outset is that EA is a long-term business strategy. On top of that, according to Ross et al. (2006) for EA to succeed, it must be sponsored at a very high level within the organization. This is because, EA is a high level corporate asset, and short term, sub-optimized benefits can be in conflict with long term enterprise wide benefits. EA provides the means to realize the benefits of investments in information systems and information technology.

Based on the theoretical works presented above it is clear that organization needs to achieve an effective EA assimilation. The importance of EA assimilation and the lack in its investigation therefore makes this study relevant.

To sum up, all prior literature reviews focused on how to improve EA optimization that focuses largely on user satisfaction assessment (van der Raadt 2011) and no empirical investigations conducted on the perspective of EA utilization. This leaves the EA community with limited information on how to promote EA utilization to achieve effective assimilation and preserved it for long term success.

## 3　Methodology

Case study research was used to accomplish the overall aim and objectives as it is characterized by its ability to get detailed information about the phenomena being investigated; TM roles and EA roles. It covers the investigation of the current state of EA, how it is being practiced and subsequently to provide insights that will help to answer the research objectives.

A government agency practicing a formal EA plan was selected for the current study. Three different individuals were interviewed using the semi-structured interview approach; one from a top level perspective and others from an employee perspective which is EA committee and IT officer to explore their EA operations and how their perception is on EA practices, in particular EA usage. The session was conducted at an approximately length of an hour and a half through face to face and on an individual basis.

The empirical data are analysed based on the EA functions on EA roles by van der Raadt et al. (2007) four dimensions were used; EA function setup, EA product, EA service delivery and EA cultural aspects. For the theory of the Gartner's activity cycle (Burton et al. 2008) for TM roles, which could help leaders to better understand their own roles. Gartner's activity cycle covers four aspects; Strategies, Architect, Lead and Govern. Later stage, the comparison between TM roles based on Gartner activity cycle and EA roles based on EA functions will result in an identification of the significant roles of TM and EA. Finally the relevant roles from the findings are synthesized which will help this study to identify how EA usage can be optimized through the involvement of TM.

# 4 Analysis and Findings

## 4.1 Enterprise Architecture Initiative in Case Study Organisation

In the government agency, the EA was initiated in 2008 anchored in the Information Management Division (IMD). The EA is regarded as a component in the IT operating mode driven by the need for integration. The committee of the EA programme was assigned to each EA domain; Business, Information, Technology and Information. The EA implemented within IMD is responsible for (1) setting the long term strategic direction and (2) reviewing all solutions developed and changes implemented by IMD.

## 4.2 Enterprise Architecture Roles

The first objective of this study is on the Enterprise Architecture (EA) roles. The identification of EA roles is based on the conceptualization of the EA functions.

### 4.2.1 EA Function setup

As the EA function setup aims at a holistic optimization of the EA in alignment with global and long term objectives, the role of this function, therefore, is to ensure that a clear EA mandate of the appointed organizational and business/IT scope is defined, central and local accountabilities as well as governance mechanism for EA decision making are defined (van der Raadt et al. 2007).

Gathered empirical data from the agency show that the current functions of EA operate on the technical level to build a common understanding for future IT direction, identify system and information needed to support business processes and document the management processes for aligning IT to business. As such, EA was perceived in the agency as having a supportive and consulting role. Not the controlling power or taking active part in initiating communication with stakeholders. The respondent expresses that they regard the involvement between EA and business process as being superfluous. The visibility of EA was not enough to make for greater usage by another division, especially business. In addition, EA were not seen as being concerned in the strategic decision.

An EA initiative in the agency is aiming at holistic optimisation of the EA central governance place a crucial role. Analysis discovered that central governance was absence from practice. Consequently, without the right degree of centralization for budget, operational process optimization and implementation, application development project prioritization and approval, IT development and implementation and infrastructure planning and management (van der Raadt 2011), EA compliance and ultimately the pursued EA goals are not enforceable.

The EA function setup also ensures the choice of EA framework and tools are in place (van der Raadt et al. 2007) guides the EA service delivery and improves efficiency and effectiveness (Bricknall et al. 2006). The EA of the agency demonstrated that having an EA framework in place guides the EA service delivery and improves efficiency and effectiveness and such framework is accepted by all relevant stakeholders as a reference for EA products.

A sub-set of the EA function is the EA principles that guide the development of an EA, as such the EA function setup ensures that EA principle provides guidance in reaching the target architecture (TOGAF). Yet in the agency, EA is not empowered to enforce the compliance of the desired IT principle of the stakeholders. EA initiative in the agency demonstrated that having a clearly defined and set up EA roles ensures that all activities are properly assigned and conducted with the right skills.

### 4.2.2 EA Products

The EA products describe the current state architecture or as-is architecture, which provides insight into the current implementation of business processes, IT systems and infrastructure; target blueprint or the to-be architecture, that will focus on the desired state in the future; roadmap from the current state to the target state, in which schedules the transformation steps (Ross et al. 1996). The first core product of EA is the documentation of the current implementation of business processes, IT systems and infrastructure. The to-be documentation describes, similar to the as-is architecture, business processes, IT systems and infrastructure, but focuses on the desired state in the future.

The transformation plan documented for the agency is a significant undertaking to schedule the transformation steps that evolve the as-is architecture step by step to the to-be architecture. It brings the transformation steps in a desired sequence accommodating contextual factors such as business priorities, budgets, and urgency (Bernard, 2012; Pulkkinen et al. 2007). Therefore the role of this dimension is to ensure that these EA artifacts are being provided with information of which characteristic and in which quality (Ross et al. 1996).

### 4.2.3　EA Service Delivery

van der Raadt (2011) described EA delivery for being responsible in providing advice to guide EA decision making, creating and maintaining EA products as well as validates solutions and operational changes to see whether they conform to the EA, and provides support in applying EA products. EA stakeholders need to be educated about the EA activities.

The problem with the agency, EA was that the information is not understandable and accessible by their stakeholders. Their EA stakeholders were not convinced of the value of EA, thus hinders their involvement and lowered the visibility of EA outside the EA function. At the same time, the EA function does not focus on a specific EA process, but on the actual services provided to external stakeholders. The EA service delivery comprises of communication, compliance validation and decision making, and support of projects where each has crucial roles (van der Raadt 2011).

To evaluate whether the set EA principles are fulfilled, regular project or architecture reviews need to be done, however this was not practiced in the agency. As such, the compliance validation and decision making were not seen as providing much support to the management in deciding on architecture and assuring project conformance. The defined approach for reviews and decision making should be transparent and consistent to be understandable by all stakeholders (Bernard, 2012; van der Raadt, 2011). In addition, top management should be briefed regarding the results of reviews and advised for decision making proactively (Ross et al. 2006).

### 4.2.4　EA Cultural Aspects

Bean (2011) claimed that the EA cultural aspect is inducted to accommodate people and soft aspects of EA. These human aspects of the EA are said to be an essential part that is often overlooked. To create an understanding for EA it is important to have a common, shared vision for the long-term as well as a common understanding of EA for the short-term, both among business and IT employees. As such, it is important that high awareness of EA be reached among all EA stakeholders. The role of this dimension, therefore, is to ensure that implicit EA values and norms are lived to implement EA successfully (Graver 2010).

The awareness of benefits from EA is not assessed to be well established in the agency. The drivers behind EA are very clear; however, it is difficult for the other departments and stakeholders to see what they are getting from EA (Rogers et al 1969). The EA initiatives at the agency failed to find resonance within the organisation due to lacking in culture of the management support. Currently EA is placed in the IT section, even though the discipline relates to the processes in all four domains. It is recommended that EA is placed in the business unit, spanning all domains, to indicate how the discipline is embedded in the model's end-to-end process chain. Therefore a common understanding of the EA must be established for both business and IT employees. It is important to have a common, shared vision for the long-term as well as a common understanding of EA for the short-term both among business and IT employees (Espinosa et al. 2011).

## 4.3　Top Management Roles

Identifying the TM role is the first step in understanding how TM can provide the support to ensure EA assimilation in order for the organization to realize its EA value (Bean 2011). As a result, it helps to identify the factors affecting the involvement of TM in supporting EA assimilation. The analysis will result in the second research objective in the identification of significant problems in the organization's EA initiative involving the TM role.

### 4.3.1　Top Management Strategize Role

In the government agency, a change of executive leadership at the top of the organization saw inoperative involvement from the TM. Findings showed that leaders were found to be less committed to EA in terms of giving EA attention, support, funding, commitment and time. The TM although was aware of the EA efforts, EA were not seen as a visible component of the top leadership's strategic and annual plans and their performance agreements.

The analysis also found that without proper oversight and support, all subsequent EA efforts are expected to fail. A key to having an effective and business-aligned EA effort is to provide an EA management and governance structure (Bernard 2012) that ensures and maintains visibility into the EA process across leadership and management functions (Niemi 2006; Ross et al. 2006). The communication is under-utilised in the government agency which was directly due to the lack of governance mechanisms. If an appropriate governance structure does not exist, then oversight boards and steering committees, with executive participation, must be implemented as the first, and highest priority for the EA effort (Buchanan 2010; Bernard 2012). In addition, this will ensure stakeholders can see from the beginning that support for the EA has buy-in from the highest levels and that structures are being put in place to ensure their voice is heard on issues.

The EA program in the agency did not include members of a major business unit within the organisation. Failing to include business units in the formulation of the management structure, the EA programme has more of an enterprise IT architecture character (refer EA Product). Not recognizing the needed contributions of all the business areas of the organization architectures in terms of the program plan, leads to the EA being overly focused on technology without adequate consideration of business function, capabilities, needs and benefits. As such, it is important that the requisite resources, training, staffing and tools for them be recognized within the consolidated EA program plan (Ross et al. 2006. pp 8-12).

### 4.3.2   Top Management Architect Role

In the government agency, during the development of EA, the TM gave a commitment on development of "as-is" and "to-be". The role also includes evaluating the current architecture view of "as-is" strategies, processes and resources and creating a management plan to move from the current view to the future. The development of the "to-be" business strategy and architecture for an organization is a significant undertaking that requires the participation and buy-in of senior business managers in an organization. Many researchers have found that this activity is often being delegated to mid-level or junior-level managers or staff in the organization, without adequate review from TM (Schekkerman, 2005). The decisions made in conducting this activity have a major impact on the effectiveness of the business transformation, the IT strategy, and the investment decisions made by the organization, and can fundamentally affect the way in which the organization will conduct business in the future. It is therefore critical to involve senior business managers in this activity from the outset and to ensure their continued oversight after the business architecture is identified.

TM championship and their sustain support throughout the phases of the project to help ensures a smooth change management (Ross et al. 2006; Todnem 2005) and mobilising commitment of other stakeholders (Purvis et al. 2001). However, in the government agency, when a change in the top leadership took place, it has affected the common vision for the target architectural state. As the case it is then realized that the vision must be modified in midstream to reflect the new reality since the "to-be" state presents a common vision for the future

### 4.3.3   Top Management Lead Role

If the communication of the services and benefits of EA is kept on a technical level, the business will be prevented from gaining a better understanding of how EA can support it (Jansen et al. 2008). Insufficient awareness was found to be an issue for EA in the government agency because the stakeholders were not found to be sufficiently communicated on the benefits that EA could give. The level of awareness is found to be lacking; either the awareness level is very low or the view on the EA function is undesirable. The government agency has not made any marketing or communication efforts to improve awareness of EA in the organization and influence the perception thereof. Communicating in technical terms does not make the benefits of EA clear about the business, but instead supports the view of EA as being a technical function. This hinders a wider acceptance of EA (van der Raat and van Vliet 2008 ). To enable a proper communication structure, a clear governance structure is needed, which indirectly results in close cooperation (van der Raadt et al. 2008).

### 4.3.4   Top Management Govern Role

EA practice in the government agency lacked of an adequate governance structure. A governance process is crucial to ensure the effectiveness of an organization's of EA (Gartner 2008; Winter and Schelp 2008). EA governance centers around creating and making sure that the EA processes and structures are followed and EA governance is thus a key aspect of ensuring positive EA performance (Weill and Ross 2003).

It was then discovered from the case study that governance initiatives and decisions are defined without a link or even knowledge of EA. Furthermore, EA committee may have been separately focused on EA processes and practices, and an engaging leader to ensure that the evolving EA reflects and supports the strategies and goals. The consequence is that diverse roles and responsibilities are often misunderstood, which could also result in significant overlaps, poor investment decisions, wasted resources and miscommunication .

From the empirical analysis, TM roles were found to be an important and strategic factor in EA practice (Schekkerman 2005). However the empirical findings discover a lack of strong executive sponsorship and ongoing leadership as a major hurdle. In brief, the most discouraging findings indicate that a strategic level in the EA practice is missing. There is very little continued involvement in the EA development by senior executives after the initial kick-off. Visible and continuous executive sponsorship permeates the rest of the organisation towards ensuring the organisation commits the right level and type of resources to conduct a successful EA practice (Armstrong and Sambamurthy 1999).

## 4.4 Conceptual Knowledge Relationship Model of Enterprise Architecture and Top Management Roles

The model draws the synthesized between EA roles with TM roles which provides high-level constructs and relationships (Figure 1).

A central component of the model is the absence of any direct path between TM support and effective EA assimilation. The link between TM support and effective EA assimilation materializes only if TM roles support EA roles. From the findings, the study hypothesize that the relationship between TM roles and EA role dimensions mediate the relationship between TM support and effective EA assimilation.

Among TM's most critical roles is strategize, it is what top managers do (Nutt, 1987). The degree of TM commitment is a critical element in shaping the EA functions setup. "One should pay special attention to the introduction strategy … and to the role of IT managers in this process" (Iivari and Huisman, 2007) as creating a foundation that centres on implementation of roles and responsibilities, defining the scope of the architecture exertion, and providing the necessary resources to effectively develop the architecture products are significant . The EA scope, EA Principles, Governance Structure and Mechanism in the EA function setup represent the elements that need to be defined by TM.

TM architect role directly affects the successful implementation of EA (Bernard, 2012) by providing desirable information about the as-is architecture, the to-be architecture and the EA roadmap or transition plan. The architect role involves (Gartner, 2007) designing the future state and articulating and base lining the current state. In addition, the gaps between the current and future states are identified, and a transition plan or road map is developed. The results "for how the organisation achieves the current and future business objectives" is referred as "blueprint" by Pereira and Sousa (2004).

EA delivery is responsible for providing advice to guide EA decision making as well as creating and maintaining EA products (Van Der Raadt 2011). The TM typically acts as the functional lead of the EA delivery function, overseeing all aspect areas of the EA. TM is responsible for the quality and effectiveness of the overall EA. The EA manager governs the EA delivery function, performing budget and resource management, planning and coordination, and other operational management tasks.

Lastly, TM lead role involves changing the culture of the organization to embrace EA, developing the architecture process corresponding with the organization's needs and capabilities, and evolving the enterprise architecture team and constituents. The degree of having executive commitment, active and sustained support from top management is crucial in shaping the EA function when the resources are more likely to be committed.

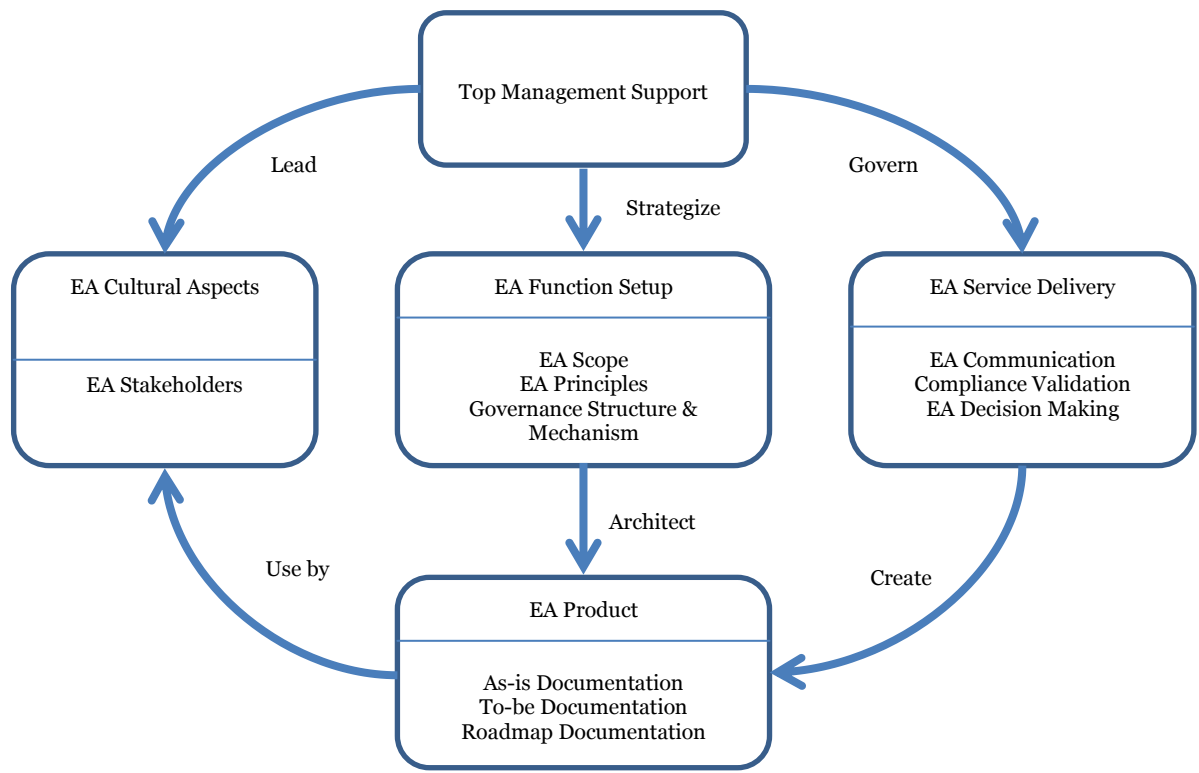

*Figure 1: A conceptual model of Top Management Roles and EA Roles Relationship*

## 5  Conclusion

The context of the case study examined the practice of EA where its main purpose was to investigate how EA is being practiced that would assist to identify issues in relation to the usage of EA.  The scope of the analysis are the areas that are linked to the role of top management and roles of EA function  to support assimilation.

The study found a lack of strong executive sponsorship and ongoing leadership as a major hurdle in EA practice. EA development is generally delegated to the Head of the IT department.  Findings from the case study had discovered that the idea of EA as still being regarded as technical function.  The EA committee respondent stated that, "EA had been the responsibility of the IT organisation … members from major business unit is not included".   This has narrowed the EA discipline significantly by both documenting and coordination of the enterprise architecture. Winter and Schelp (2008) emphasis that " business units have to be integrated into all EA management processes"  even when EA is being utilized for communicating business requirements to the IT units only.

Based on the findings, it can be learned that organizations must reorganize the many aspects of EA that must be managed to ensure the EA initiative is an ongoing success to the organisation. Findings also discover the organization must not only focus in documenting the current state or future state, or creating principles and models of EA but to invest in all the aspects of EA studies. The high level of IS and business knowledge areas relevant for TM to support the EA usage, namely knowledge of the business and knowledge of the technology (Cooprider and Victor 1993).

EA research that addresses post adaptive EA utilization is limited (Van Der Raadt, 2011). Thus, this study is an early contributor  to this research stream, particularly in the context of Malaysia. On top of the study's contributions to the existing EA body of knowledge, on the importance of the knowledge needed for the senior management to support not just the usage of EA but also planning and implementing as well, this study also extends the understanding of EA utilization specifically on the kinds of knowledge for TM involvement that best supports understanding. This could give significant insights into some areas that require improvement or areas that needs further development. Researchers may use the knowledge to establish factors that can influence progress in assimilation.

The last contribution of this study is for EA practitioners. Leveraging on the EA knowledge can be potential in improving the current practice of practitioners that will translate into tangible and intangible benefits for organizations. The conceptual relationship model presented can be used as guidelines for managers to find ways to increase the correct use of the system by its users, while taking into consideration the capabilities of the system, the organizational needs and the context where the EA system is being used.

## 6   Limitations and Suggestions for Future Research

Most studies have certain limitations. Same goes to this study. The limitations and suggestions for future studies are as follows: firstly, this study focuses on developing a universally applicable EA knowledge that most EA practitioners should have, to be able to ensure EA usage is optimized in achieving effective EA assimilation. Therefore, future research can cover in-depth the methodology and principle, e.g. EA framework, used in designing and maintaining EA. Secondly, the content of this study on general TM roles. In-depth study could be done on the specific role of TM in assimilating EA. Example during the budget preparation. Thirdly, the semi-structured interview was the only method used to collect the required data. Hence, in order to better understand the views in more details on the roles of TM and EA, further research involving in-depth interviews or focus groups. Fourthly, the respondent was from only one organization which is a public sector. Therefore, future research should include more respondents from other organization that have experience and also to include an expert from other industries. Lastly, this study just proposed the conceptual model as a guideline. It is suggested that the model should be tested for future research. Despite the limitations, this study provides useful insight to the TM on the EA knowledge needed to ensure the usage of EA for the benefits of the organization.

## References

Akhouri, N. 2002. *Aligning HRM Strategies with Competitiveness*. Singh, S (Ed.), High Performance Organization. New Delhi, New Age.

Amour, F.J., Kaisler, S.H., and Liu, S.Y. 1999. "Building an Enterprise Architecture Step by Step," *IT Professional (*1), pp 31-39.

Armstrong, C.P., and Sambamurthy, V. 1999. "Information Technology Assimilation in Firms," *Information Systems Research (*10:4), pp 304-327.

Aziz, S., Obitz, T., Modi, R., and Sarkar, S. 2005. Enterprise Architecture: A Governance Framework http://www.infosys.com/consulting/architecture-services/white-papers/Documents/EA-governance-1.pdf. Retrieved: 16 April 2015.

Bean, S., 2011. Re-Thinking Enterprise Architecture Using Systems And Complexity Approaches, http://www.irmuk.co.uk/articles/s_bean_re_thinking_enterprise_architecture.pdf. Retrieved: 28 April 2015.

Bernard, S. A. 2012. *An introduction to enterprise architecture*. AuthorHouse.

Broers, C.M. 2005. Career and Family: The Role of Social Support, Griffith University, PhD Thesis.

Bricknall, R., Darrell, G., Nilsson, H., and Pessi, K. 2006. Enterprise Architecture: Critical Factors Affecting Modelling and Management. *ECIS 2006 Proceedings*. Paper 146.

Chatterjee, D.D., Pacini, C., and Sambamurthy, V. 2002. "The Shareholder-Wealth and Trading-Volume Effects of Information Technology Infrastructure Investments," *Journal of Management Information Systems* (19:2), pp 7-42.

Dankova, P. 2009. "Main Aspects of Enterprise Architecture Concept," *Economics Alternatives*, Issue 1.

DeLone, W. H., and McLean, E. R. 2002. Information Systems Success Revisited. In *System Sciences, 2002. HICSS. Proceedings of the 35th Annual Hawaii International Conference on System Sciences*. pp 2966-2976. IEEE.

Espinosa, J. A., Boh, W. F., and DeLone, W. 2011. The Organizational Impact of Enterprise Architecture: A Research Framework. In *System Sciences (HICSS), 2011 44th Hawaii International Conference on System Sciences*. pp 1-10. IEEE.

Fichman, R.G. 2000. *The Diffusion and Assimilation of Information Technology Innovations*. R.W. Zmud (Ed.), *Framing the Domains of IT Management: Projecting the Future Through the Past*, Pinnaflex Educational Resources, Inc., Cincinnatti, OH.

Fichman, R.G., and Kemerer, C.F. 1999. "The Illusory Diffusion of Innovation: An Examination of Assimilation Gaps," *Information Systems Research* (10:3), pp 255-275.

Gaver, S. 2010. An Examination Why the Federal Enterprise Architecture Program Has Not Delivered the Expected Results and What Can be Done About It. *Why Doesn't the Federal Enterprise Architecture Work?*

Hambrick, D. 2007. Upper Echelon Theory: Revisited. *Academy of Management Review (32:2)*, pp 343.

Hambrick, D., and Mason, P. 1984. "Upper Echelons: The Organization as a Reflection of its Top Managers," *Academy of Management Review (9:2), pp 193-206.*

Hite, R.C., Barkakati, N., Bird, M., Collier, B., Davis, D., Doherty, N., Goldstein, T., and Tekeley, R. 2003. "Information Technology: A Framework for Assessing and Improving Enterprise Architecture Management," White Paper, United States General Accounting Office, Washington, DC.

Iivari, J., and Huisman, M. 2007. The relationship between organizational culture and the deployment of systems development methodologies. *Mis Quarterly*, 35-58.

Jasperson, J.S., Carter, P.E., and Zmud, R.W. 2005. "A Comprehensive Conceptualization of Postadoptive Behaviors Associated with Information Technology Enabled Work Systems," *MIS Quarterly* (29:3), pp 525-557.

Jansen, A., de Vries, T., Avgeriou, P., & van Veelen, M. 2008. *Sharing the Architectural Knowledge of Quantitative Analysis*. In Quality of Software Architectures. Models and Architectures. Springer Berlin Heidelberg.

Jensen, S. 2010. *Government Enterprise Architecture Adoption: A Systemic-Discursive Critique and Reconceptualisation*. IT University, Copenhage, Denmark

Kappelman, L. 2009. *The SIM Guide to Enterprise Architecture: Creating the Information Age Enterprise*. New York: Auerbach.

Lange. M. and Mendling, J. 2011. An Experts' Perspective on Enterprise Architecture Goals, Framework Adoption and Benefit Assessment. *Proceedings of the 6th Trends in Enterprise Architecture Research Workshop (EDOCW'11)*.

Lankhorst, M. 2005. *Enterprise Architecture at Work: Modelling, Communication and Analysis*, Springer.

Lemmetti, J. and Pekkola, S. 2012. "Understanding Enterprise Architecture: Perceptions by the Finnish Public Sector," *Electronic Government*, pp 162-173.

Liang, Z., and Zhu, X. 2007. "MNCs' Patent Strategy in China and its Implications", *China Soft Science*, No.1, pp 55-61.

Makiya, G.K., and Lyytinen, K. 2011. *The Antecedents of Enterprise Architecture Assimilation in Organizations*. Unpublished Quantitative Research Report, Weatherhead School of Management, Case Western Reserve University.

Markus, M.L., and Tanis, C. 2000. "The Enterprise System Experience - from Adoption to Success", in Zmud, R.W. (Ed.), *Framing the Domains of IT Management: Projecting the Future Through the Past*, Pinnaflex Educational Resources, Inc., Cincinnatti, OH, pp. 173-207.

Neimi, E. 2006. Enterprise Architecture Benefits: Perceptions from Literature and Practice. *The Proceedings of the 7th IBIMA Conference Internet & Information Systems in the Digital Age*, 14 16 December, Brescia, Italy.

Nutt, P. C. 1987. Identifying and appraising how managers install strategy.*Strategic Management Journal* (*8:*1), pp 1-14.

Perko, J. 2008. IT Governance and Enterprise Architecture as Prerequisites for Assimilation of Service-Oriented Architecture. *Doctoral Dissertation*, http://dspace.cc.tut.fi/dpub/handle/123456789/151. Retrieved: 16 April. 2015

Pulkkinen, J. 2007. Cultural globalization and integration of ICT in education. In K. Kumpulainen (Ed.), *Educational technology: Opportunities and challenges,* Oulu, Finland, University of Oulu, pp. 13–23.

Purvis, R.L., Sambamurthy, V., and Zmud, R. W. 2001. "The Assimilation of Knowledge Platforms in Organizations: An Empirical Investigation," *Organization Science (*12:2), pp 117–135.

Rafidah, A.R., Zulkhairi, M.D., Rohaya, D., Siti-Sakira, K., and Sahadah, A. 2007. "Enterprise Information Architecture (EIA).: Assessment of Current Practices in Malaysian Organizations," 40th Annual *Hawaii International Conference on System Sciences,* Hawaii, IEEE pp 219a-219a.

Robbins, S.P. 1998. *Organizational Behavior: Concepts, Controversies, Applications.* Upper Saddle River, NJ: Prentice-Hall.

Ross, J. W., Beath, C., and Goodhue, D. L. (1996) Develop long-term competitiveness through IT assets. *Sloan Management Review (* 38), pp 31-45.

Ross, J.W., Weill, P., and Robertson, D. 2006. *Enterprise Architecture As Strategy: Creating a Foundation for Business Execution*, Harvard Business School Press.

Saha, P. 2006. "A Real Options Perspective To Enterprise Architecture as an Investment Activity," *Journal of Enterprise Architecture (*2)**,** pp 32.

Schekkerman, J. 2005. Report of the Third Measurement December 2005. *Trends in Enterprise Architecture*. Institute For Enterprise Architecture Developments.

Seppänen, V., Heikkilä, J., and Liimatainen, K., 2009. "Key Issues in EAimplementation: Case study of Two Finnish Government Agencies," Proceedings of the 11th IEEE Conference on Commerce and Enterprise Computing, IEEE, pp 114-120.

Somers, T.M., and Nelson, K.G. 2004. "A Taxonomy of Players and Activities Across the ERP Project Life Cycle," *Information & Management* (41), pp. 257-78.

Tamm, T., Seddon, P.B., Shanks, G., and Reynolds, P. 2011. "How Does Enterprise Architecture Add Value to Organisations?," *Communications of the Association for Information Systems (*28:1), pp 141-168.

TOGAF—The Open Group Architectural Framework (2005). http://www.togaf.org. Retrieved: 12 September, 2014.

Tucker, C. and Aron, D. 2005. Executive summary "Applying Enterprise Architecture," , http://www.marketwired.com/press-release/provision-50-combines-enterprise-architecture-with-bpm-complete-enterprise-modeling-735578.htm Retrieved: 16 June 2014.

van der Raadt, B. 2011. Enterprise Architecture Coming of Age. *SIKS Dissertation,* Series No. 2011-5.

van der Raadt, B., Slot, R., and van Vliet, H. 2007. Experience Report: Assessing a Global Financial Services Company on its Enterprise Architecture Effectiveness Using NAOMI. *Proceedings of the 40th Annual Hawaii International Conference on System Sciences.*

van der Raadt, B., Schouten, S., and van Vliet, H. 2008. *Stakeholder Perception of Enterprise Architecture. In Software Architecture*. Springer Berlin Heidelberg.

van der Raadt, B., and Van Vliet, H. 2008. *Designing the Enterprise Architecture Function. In Quality of Software Architectures. Models and Architectures*. Springer Berlin Heidelberg.

Weill, P. and Ross, J.W. 2003. *IT Governance – How Top Performers Manage IT Decision Rights for Superior Results*. Boston: Harvard Business School Press.

Winter, R. and Schelp, J. 2008. Enterprise Architecture Governance: The Need for a Business to IT Approach

Zachman, J.A., 1987. "A Framework for Information Systems Architecture," *IBM Systems Journal (*38:(2/3), pp 454-470.

Zachman, J.A. 2003. The Zachman Framework: A Primer for Enterprise Engineering and Manufacturing, http://www.businessrulesgroup.org/BRWG_RFI/ZachmanBookRFIextract.pdf Retrieved: 24 May. 2014.

Zachman, J.A., and. Sowa, J.F. 1992. "Extending and Formalizing the Framework for Information Systems Architecture," *IBM Systems Journal* (31:3), pp 590-616.

Zulkhairi, M.D., Rafidah, A.R., Rohaya, D., Siti-Sakira, K., and Sahadah, A. 2006. "Enterprise Information Architecture: Empirical Evidence to Support Zachman Framework in Malaysia", Knowledge Management International Conference and Exhibition (KMICE'06), Kuala Lumpur, Malaysia, 6-8 June, 2006.

## Acknowledgements

This study was conducted as a part of a two year research project focusing on enterprise architecture knowledge of top management. It is coordinated by the Research Management Centre (RMC) in International Islamic University Malaysia (IIUM), and funded by the Malaysian government under the Ministry of Education (MOE). We wish to thank the government agency for their assistance in providing us the valuable data.